  \documentclass[multi]{cambridge6Atight}
  \usepackage{natbib}

  \usepackage{rotating}
  \usepackage{floatpag}
  \rotfloatpagestyle{empty}

  \usepackage{amsthm}
  \usepackage{graphicx}
  \usepackage{mathptmx}

  \usepackage{makeidx}
  \makeindex

 \copyrightline{This material has been published in \textit{The Impact of Binaries on Stellar Evolution}, Beccari G. \& Boffin H.M.J. (Eds.). This version is free to view and download for personal use only. Not for re-distribution, re-sale or use in derivative works \copyright\ 2018 Cambridge
    University Press.}
    
\begin{document}


  \alphafootnotes
   \author[J.\,J.\, Eldridge \& E.\,R.\, Stanway]
    {J.J. Eldridge\footnotemark\
    and Elizabeth R. Stanway}
  \chapter[Population and spectral synthesis: it doesn’t work without binaries]{Population and spectral synthesis: it doesn’t work without binaries}


  \contributor{J.J. Eldridge
    \affiliation{Department of Physics, University of Auckland,
 Private Bag 92019, Auckland, New Zealand. }}

  \contributor{Elizabeth R. Stanway
    \affiliation{Department of Physics, University of Warwick,
 Gibbet Hill Road, Coventry, CV4 7AL, UK.}}

 \begin{abstract}
In this chapter we discuss the population and spectral synthesis of
stellar populations. We describe the method required to achieve such
synthesis and discuss examples where inclusion of interacting binaries
are vital to reproducing the properties of observed stellar
systems. These examples include the Hertzsprung-Russell diagram,
massive star number counts, core-collapse supernovae and the ionising
radiation from stellar populations that power both nearby HII regions
and the epoch of reionization. We finally offer some speculations on
the future paths of research in spectral synthesis.
 \end{abstract}

\section{What is population and spectral synthesis?}

Population synthesis involves predicting the parameters of a
population of stars through a combination of individual stellar models
or empirical templates. Examples of stellar populations include those
in individual star clusters, galaxies or those that give rise to a
sample of transient events such as core-collapse supernovae or
gravitational wave merger `chirps'. To make such predictions through
population synthesis a number of stellar models, each predicting the
properties and evolution of one star, are combined together with an
initial parameter distribution to calculate a synthetic population
that can be compared to observational constraints including stellar
type ratios and transient event rates.

Spectral synthesis is the term for the combination of such a synthetic
population with matched stellar atmosphere models that predict how
each star of known surface temperature, gravity and composition would
appear if observed across all wavelengths. The resulting composite
spectral energy distribution can be convolved with the known
parameters of individual detectors and telescopes. This step
effectively observes the model population so it can be directly
compared to the observational data. This technique is especially
important when attempting to interpret unresolved stellar populations
where individual stars cannot be studied and classified individually.

The first population synthesis models can be traced back to the first
attempts to understand the stellar content of galaxies and star
clusters \citep[e.g.][]{1968ApJ...151..547T,1976ApJ...203...52T} There
are now many codes that perform population and spectral synthesis. The
evolutionary models most widely used to date carry the implicit
assumption that most stars evolve as single stars, without any
significant binary interaction, and thus relatively few evolutionary
avenues are available
\citep[e.g.][]{1999ApJS..123....3L,2003MNRAS.344.1000B,2016ApJ...823..102C}. While
these codes have allowed us to deepen our knowledge of stellar
systems, they are limited by their single-star evolutionary
paradigm. This is fundamentally an unphysical assumption. It is clear
from other chapters in this book that the fraction of binary stars in
stellar populations range from 20\% for the lowest mass stars to 100\%
for more massive stars.  As this volume also indicates, in many
binaries at least one of the stars will have a substantially different
evolutionary path to that of a single star. As a result, a failure to
account for binary effects would be expected to compromise
interpretation of stellar system parameters in any regime where these
are significant for the dominant population.

Over the past few decades a number of groups have been attempting to
account for interacting binaries in their population synthesis, and
more rarely spectral synthesis. Groups computing population synthesis
models include, for example,
\citet{1996MNRAS.280.1035T,2002MNRAS.329..897H,2004A&A...419.1057W,2004NewAR..48..861D,2005MNRAS.364..503Z,2007ApJ...662..504B,2009ARep...53..915L,2009A&A...508.1359I,2013A&A...557A..87T,2017PASA...34...58E}. However
from this list only \citet{2004NewAR..48..861D,2005MNRAS.364..503Z}
and \citet{2017PASA...34...58E} also make spectral synthesis
predictions. Inclusion of spectral synthesis within binary population
synthesis models has confirmed that interacting binaries make a
substantial change to the predictions of spectral synthesis and that
the use of single star models can leads to errors in interpreting
certain aspects of stellar populations.

In this chapter we first give a brief overview of how population and
spectral synthesis is performed and highlight some of the key
uncertainties. We then discuss selected example observations where
interacting binaries make a substantial difference to how we
understand the underlying stellar systems. These include the HR
diagram of resolved stellar clusters, number counts of massive stars
in galaxies, core-collapse supernovae, gravitational wave sources, the
stellar populations of distant galaxies and the epoch of reionization.

\section{How do you do it?}

To perform a population synthesis one first has to gather or create a
set of stellar evolution models from which to construct the synthetic
population. It is possible either to use extant stellar
models \citep[e.g.][]{2012A&A...537A.146E} or to calculate your own
\citep[e.g.][]{2016ApJ...823..102C,2017PASA...34...58E}. It should be
noted that these models themselves incorporate substantial
uncertainties. There are many known problems in stellar evolution such
as determination of mass-loss rates, rotation rates and how to
implement convection that all impact on the certainty with which
predictions of stellar properties at a given mass and age can be
made. We can however test these models by comparing their properties
to observed individual stars such as the Sun or the physically
well-constrained main-sequence stars in eclipsing binaries.

When computing models of interacting binary stars the situation is
more complicated due to the extra physics that needs to be included in
evolution codes. Prescriptions for Roche lobe overflow (RLOF) and
common envelope evolution (CEE) must be included. An extensive body of
literature exists exploring the uncertainties on each of these and the
tunable parameters in any given prescription. Nonetheless, these are
vital to account for the mass transfer from one star to the other in a
binary, and to shrink the orbit to produce the tight post-mass
transfer binaries we see in the Galaxy. The effects of any mass
transfer on the subsequent evolution must then be considered.

There are two different approaches to modelling the evolution of
binary stars. One of these is to use a detailed stellar evolution code
\citep[e.g.][]{2004NewAR..48..861D,2017PASA...34...58E} however this
has the drawback that each model can take several minutes to
calculate. This means that to calculate a large grid of models in
order to full investigate the parameter space can take years on a
single computer. With Universities investing in computational
resources and large clusters it is beginning to become feasible to use
detailed models to create the tens of thousands of models required for
population synthesis.

Alternately approximate rapid codes, for example
\citet{2002MNRAS.329..897H}, approximate the evolution of binary stars
by fitting functions to a smaller pre-calculated grid of detailed
models and interpolating between them. While the evolution is only
followed in an approximate manner, this method does allow multiple
large grids to be rapidly calculated so that how uncertainties in the
input physics effect the results of the population synthesis can be
investigated, \citep[e.g.][]{2013ApJ...764..166D}. Applications
include exploring the effects of stellar rotation on evolutionary
stages. Such a study would be challenging in a detailed code as the
each model run would need to follow the transport of angular momentum
through the star in detail, which makes the models difficult (and also
time consuming) to calculate \citep[e.g.][]{2007A&A...465L..29C}.

Once the individual stellar models are calculated they must be
combined into a synthetic population. This is done by weighing each
model by a factor that represents how likely it is to exist in a given
scenario. For all populations we use an initial mass function that
determines how likely it is that a star of a given stellar mass exists
at a zero main-sequence age after the initial star formation
episode. Examples include \citet{1993MNRAS.262..545K} and
\citet{2003PASP..115..763C}. These all are more modern versions of the
model established by Salpeter [1955] where a power law representation
was used to capture the observed fact that there are many more low
mass than high mass stars.

In a binary population we must also describe the distribution of
initial periods, eccentricities and secondary masses. Some of the most
recent can be found in \citet{2017ApJS..230...15M}. The binary
fraction is around 100\% of stars being in binaries for stars
$\ge10$\,M$_{\odot}$ with 70\% of them close enough to have their
evolution affected by binary interactions during their evolution
\citep{2012Sci...337..444S}. For less massive stars, less massive than
$< 1$\,M$_{\odot}$, the binary fraction drops to 20 to 40\%.

The period distributions also vary with stellar mass. While Opik's law is a good
first estimate, studies tend to indicate that
there may be more close binaries. Nonetheless a flat distribution in the mass
ratio for the secondary seems a good approximation. More data is still
required to gain a firm understanding, but the initial parameters of
binaries are beginning to become well constrained.

To summarise we now know how to weight each of our binary models and
have a model that follows the evolution of the stars, but there is one
final complication: the first supernova. When a supernova occurs in a
binary there are one of two possible outcomes, either the binary is
unbound and the two stars go on to evolve as single stars or they
remain bound and become a binary with a compact companion. These may
evolve into a X-ray binary.

Determining whether the system is unbound or not depends on a few
factors. First is the amount of ejecta in the supernova. It can be
shown, assuming a circular pre-SN orbit, that if the ejecta mass is
more than half the mass of the binary then the system is unbound. Therefore
estimating how much mass goes into the remnant and how much is ejected
is a key problem. Work such as \citet{2016ApJ...821...38S} is showing that it may not
be easy to determine this accurately. The final outcome of evolution
appears to be chaotic over some mass ranges and depends on all the
uncertainties put into the stellar model. However some simpler schemes
exist to estimate the remnant mass and future work will allow us to understand
this more.

The other factor is that in formation neutron stars and black holes
appear to be given a momentum kick during core collapse which can
range from a few 10s km/s up to 1000km/s!
\citep[e.g.][]{2005MNRAS.360..974H}. The source of the kick is
currently being investigated by many groups
\citep[e.g.][]{2016MNRAS.461.3747B,2017ApJ...837...84J}.  It is
important to consider this as kicks can change the orbital velocity of
the compact remnant in the SN and this can lead to unbinding binaries
that would have remained bound as well as keeping binaries bound that
would otherwise have been unbound. Further complicating the issue is
whether black hole have kicks or not; most prescriptions seem to
indicate they must be weaker but there is limited observational
evidence \citep[e.g][]{2016MNRAS.456..578M}.

In our population synthesis the first supernova therefore creates
multiple possible future evolutionary pathways for each binary. These
have to be calculated and are usually accounted for by Monte Carlo
methods, making models with a large number of different random kicks
in direction and magnitude and weighting the results by a probability
distribution. This is again where the rapid models have an
advantage as many different models can be calculated quickly, while for
detailed models some approximations must be made to calculate such
stellar models in a reasonable timeframe.

Once a synthetic population is created it can be compared to observed
stellar populations. For observations such as number counts,
distributions of luminosities or orbital period distributions this can
be fairly straightforward. The stumbling block is that typically the
observations might only be in a few observed photometric filters, or
comprise only the optical spectrum of a star. Most stellar models only
give a bolometric luminosity and effective temperature for the stars,
and so linking this to a spectrum or photometric V band magnitude
requires an additional step.

It is possible to take the observations and process them, accounting
for their selection functions, to get the required numbers to compare
to the models. This is some what risky when we have an incomplete data
set where there are large gaps in our knowledge. A more rigorous
method is to take the stellar models and observe the models in a
similar way. This can be done by attaching stellar atmosphere models
to the stellar evolution models and creating a synthetic
spectrum. Doing so requires bringing together a large number of
different spectra from different sources as the physics of cool and
hot stellar atmospheres can be very different and the best models for
each regime may be produced by different teams and codes
\citep[e.g.][]{2002A&A...381..524W,2002MNRAS.337.1309S,2006A&A...457.1015H,2008A&A...486..951G,2012A&A...540A.144S,2013ApJ...768....6M}

Matching the stellar model to the atmosphere model typically requires us to
know the effective temperature, gravity and surface composition of our stellar model as
these are the primary factors that determine the stellar spectra. Once
this match is done it is possible to create either spectra for
individual stars or for the entire population, allowing us to closely
match the observations to the models and gain greater constraints even
if the observational dataset is limited.

All these complications and uncertainties have in the past contributed
to an air of distrust regarding binary population synthesis. Today,
however, the number of firm constraints on the uncertain parameters is
increasing. Single model populations are being compared to a large
number of varied observables, as performed by the BPASS team in
\citet{2017PASA...34...58E}, giving additional confidence in their
application. Other groups such as the Brussels group
\citep[e.g.][]{1999NewA....4..173V,2000A&A...358..462V,2007ApJ...662L.107V},
binary\_c community
\citep[e.g.][]{2009A&A...508.1359I,2014A&A...563A..83C,2015A&A...581A..62A}
and Yunnan groups
\citep[e.g.][]{2005MNRAS.364..503Z,2007MNRAS.380.1098H,2014ApJS..215....2H,2015MNRAS.447L..21Z}
are also doing very similar work.

The key result emerging from all these studies however is that
we MUST include interacting binaries in our population and spectral
synthesis. If we do not then we run the risk of drawing incorrect
conclusions when studying systems using stellar population models as our
tool.

\section{Why are binaries important?}

Binaries are important as they provide evolutionary pathways that are
simply not accessible from single-star evolution. Single star models
have for some time done a very good job of allowing us to understand
stellar populations which, in the local Universe, are often old and
metal rich and thus relatively unaffected by binary
interactions. Recent observational evidence shows that stars are born
with many of them in binaries, especially the most massive
\citep[e.g.][]{2012Sci...337..444S,2017ApJS..230...15M} and so the
effects on young stellar populations is likely to be more
pronounced. As well as that, many stars stay in a binary system right
to the last possible event in a binary's life, when the two remnants
of the stars inspiral and merge due to the loss of orbital energy
through gravitational radiation
\citep[e.g.][]{2016PhRvL.116x1102A,2017ApJ...848L..12A}.

This suggests we need to take a holistic view when attempting to model
stellar populations. The effect of binaries will not always be obvious
and may only become apparent in certain phases of evolution. For
example, the main sequence evolution is not significantly affected by
binary interactions (although mergers do occur and there are a few exceptional
cases) when looking at a population. Most differences between
predictions from single stars and binaries become apparent in
post-main sequence phases of evolution and the eventual deaths-throes
of massive stars in core-collapse supernovae.

Rather than concentrate on one detail of population and spectral
synthesis predicts we will introduce a varied number of observations
and show that in each case interacting binaries are required to
explain the observed systems, using our own BPASS models [Eldridge et
  al 2017] as a demonstration. We will first consider stellar
populations, then consider spectral synthesis before finally
speculating about the future directions of what population synthesis
will be able to predict in future.

\subsection{Blue stragglers on the HR diagram}

For resolved stellar populations in clusters the effects of binary
evolution are two-fold. There are two main binary interactions that
affect the position of stars on a Hertzsprung-Russell diagram for
example: mergers and mass transfer. These make more massive
stars available at ages for a cluster well beyond their expected
main-sequence turn off. They are well known in globular clusters for
example as blue stragglers, for younger H\,II regions or open clusters
identifying such stars is more difficult.

\begin{figure}
\begin{center}
\includegraphics[width=0.95\textwidth]{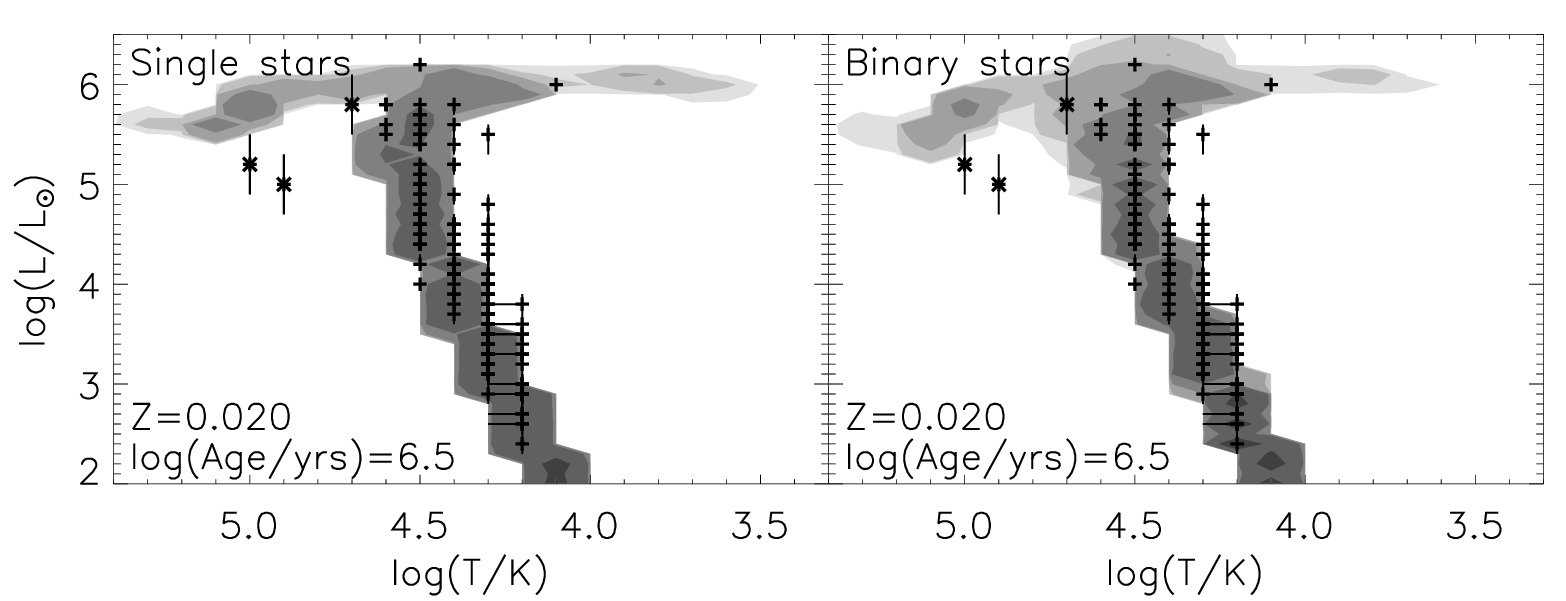}
\includegraphics[width=0.95\textwidth]{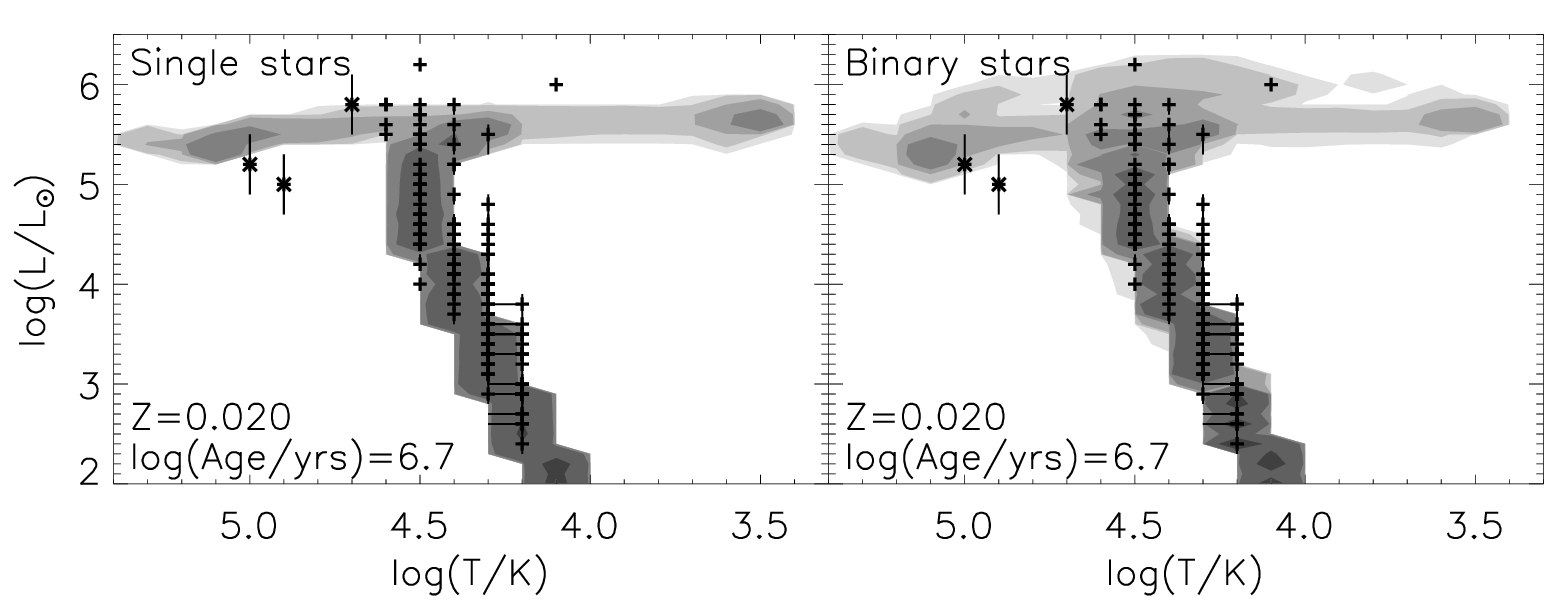}
\end{center}
\caption{The HR diagram of Cygnus OB2
  \citep{2012ApJ...751....4K,2015MNRAS.449..741W} compared to
  theoretical HR contours for single star (left panels) and binary
  star (right panels) populations. The models are for two different
  ages, 3Myrs in the upper panels and 5 Myrs in the lower
  panels.}.\label{fig:bluestragglers}
\end{figure}

For clusters the age is frequently estimated by using single star
isochrones, primarily from attempting to fit the main-sequence turn
off. The existence of blue stragglers can directly affect the accuracy
of this method. We show in Figure \ref{fig:bluestragglers} an example
fit to a single cluster, with single-star isochrones and a population
including interacting binaries, which yield an isochronal contour
plot. We see that for the well known cluster Cygnus OB, to fit the
most luminous stars a maximum age of 3 Myrs is required from single
stars, but with the binary population an older age of 5 Myrs is
qualitatively a better match to the data. While this is an extreme
example it shows that there is something that must be taken into
consideration when attempting to evaluate the age of resolved
clusters.

\subsection{Number counts}

\begin{figure}
\begin{center}
\includegraphics[width=0.8\textwidth]{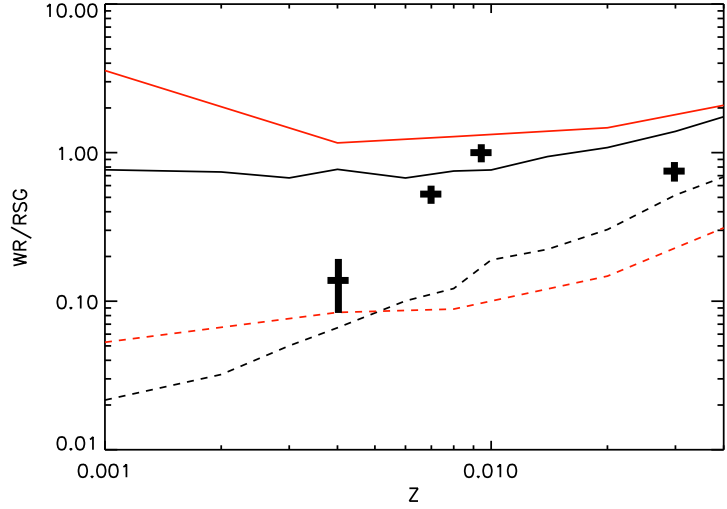}
\end{center}
\caption{The ratio of the number of Wolf-Rayet stars to red
  supergiants as observed in galaxies at a single metallicity. Solid
  lines show BPASS binary models and dashed lines show single models.
  Points with error bars show observational constraints taken from the
  literature. All ratios include only stars with $\log(L/L_{\odot}) >
  4.9$. The red lines show results for constant star formation from
  BPASS v1.0, while the black lines show similar results from BPASS
  v2.1 with IMF $M_{\rm max} = 100 M_{\odot}$, respectively). The
  WR/RSG observed ratios come from P. Massey (private
  communication).}.\label{fig:numbercounts}
\end{figure}

Another basic observable of stellar populations are number
counts. If stars can be typed then, say, the total number of O stars
can be compared to the number of red supergiants (RSG) or Wolf-Rayet (WR)
stars. This is a measurement of the number of main-sequence to
post-main sequence stars: something that depends on the nuclear
evolution of a star’s core as well as the mass loss from the
surface. For example in Figure \ref{fig:numbercounts} we show an
example comparing the observed number of red supergiants to Wolf-Rayet
stars, post-main sequence stars that have retained or lost their
hydrogen envelopes respectively. We can see here the dramatic change
in the ratio of these two stellar types that arises due to the
inclusion of binary interactions. The change is roughly around an
order of magnitude due to the binary interactions providing an extra
avenue for mass loss rather than stellar winds alone. The binary
models appear to match the observations slightly better, although the
lowest metallicity point in the SMC is closer to the single star
evolution line. However the assumed scaling of mass-loss rates might
be too strong or the assumption of constant star formation may also be
wrong with the plot only based on $\approx$100 RSGs and 13 WRs.

\subsection{SN progenitors}

If the effect of interacting binaries is so dramatic that we can infer
its presence from the number ratios of stars, we should also find this
reflected in the number of stars that die with their hydrogen
envelopes intact or removed. When core-collapse supernovae occur they
have historically been classified by their observational
characteristics, primarily their spectra and lightcurves. The broadest
classification is made on whether they are hydrogen-rich, type II, or
hydrogen-free, type Ib/c (note type Ia supernovae arise from
thermonuclear detonations of carbon-oxygen white dwarfs and are
identified by strong silicon lines).

\begin{figure}
\begin{center}
\includegraphics[width=0.8\textwidth]{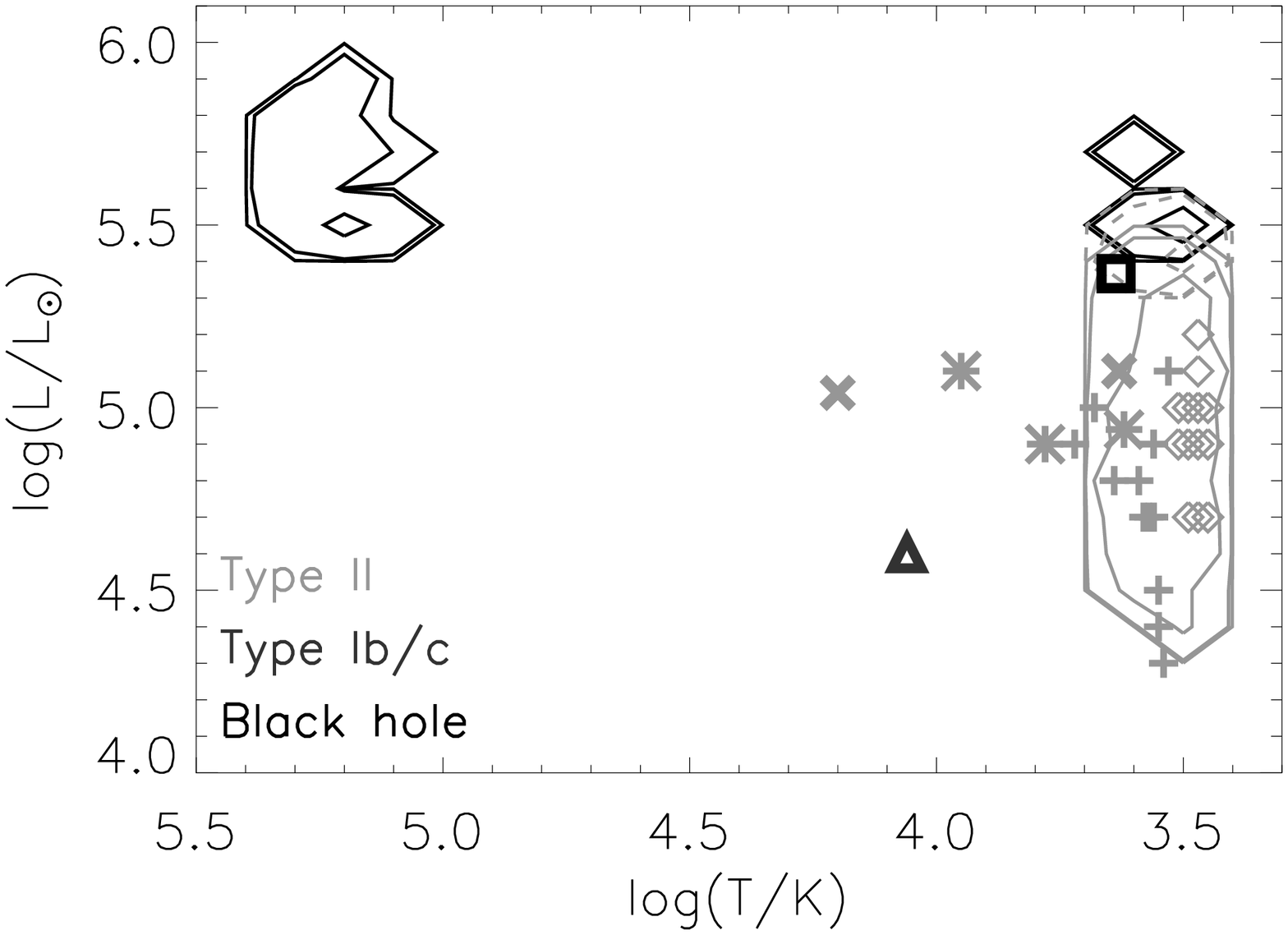}
\includegraphics[width=0.8\textwidth]{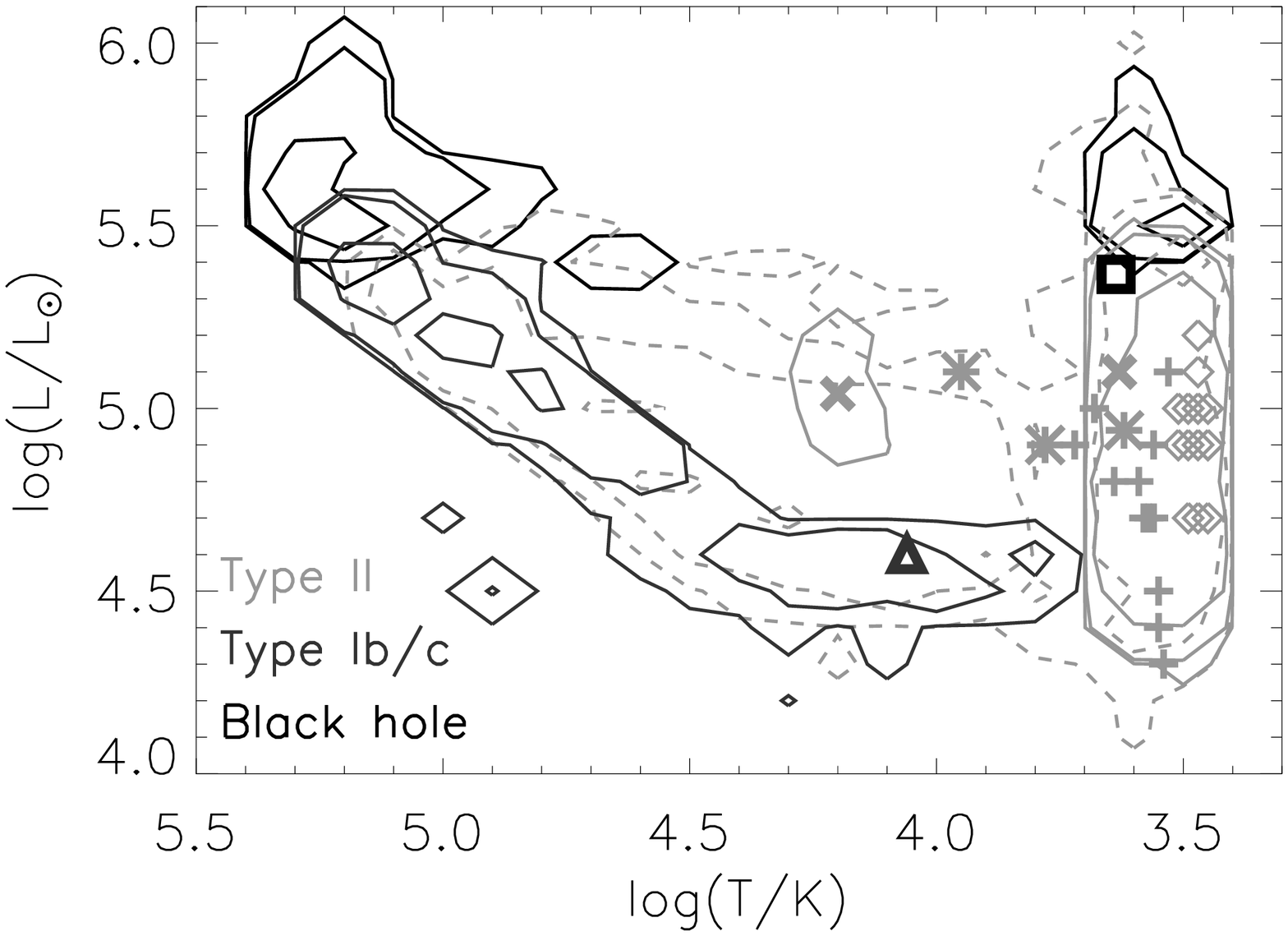}
\end{center}
\caption{HR diagrams showing the predicted location of SN progenitors
  from BPASS v2.1 models compared to the position of observed SN
  progenitors. The upper panel is for single star populations and the
  lower panel is for binary star populations. The black contours are
  for progenitors expected to form black holes, the light grey
  contours are for type II progenitors and the dark grey contours are
  for type Ib/c progenitors. The points are observations taken from
  \citet{2015PASA...32...16S}, the plus' are observed type IIP
  progenitors, the diamonds are type IIP progenitors with upper limits
  on the luminosity, the asterisks are type IIb progenitors, the
  crosses are the progenitors of 1987A and 1993J, the triangle is for
  the type Ib progenitor of iPTF13bvn and the square is for the
  candidate black-hole forming event from
  \citet{2017MNRAS.469.1445A}.}\label{fig:progenitors}
\end{figure}

As we show in Figure \ref{fig:progenitors}, the type II supernovae
progenitors are known to be mostly red supergiants with some blue and
yellow supergiants \citep[see][]{2015PASA...32...16S}. For type Ib/c
supernovae the progenitors were long expected to be Wolf-Rayet stars
but a problem with this was that if this was the case then they should
have been directly detected in pre-explosion images of a sample of
nearby explosions. This suggested that the progenitors were more
likely to be lower-mass helium stars that were the result of binary
interactions
\citep{2012A&A...544L..11Y,2013MNRAS.436..774E}. Confirmation that
this is likely to be the case came from observations of the progenitor
of type Ib supernova, iPTF13bvn. Detailed modelling by many groups
have shown that the most likely progenitor of this event is a binary
system
\citep{2014AJ....148...68B,2016MNRAS.461L.117E,2017ApJ...840...10Y}.
We can see this in Figure \ref{fig:progenitors} as in single star
predictions progenitors can only either be cool red supergiants or hot
Wolf-Rayet stars. The binary predictions however fill a much greater
space of the HR diagram, especially filling the region where most
progenitors have been observed to date.

However while progenitor detections provide direct evidence for the
effects of binary evolution, the most compelling evidence comes from
the relative rate of type Ib/c to type II progenitors. A consistent
model should be able to predict both the WR/RSG ratio as well as the
type Ib/c to II relative rate. The ratio is dependent on metallicity,
with fewer type Ib/c supernovae at lower metallicities, but as shown
by the recent study of \citet{2017ApJ...837..120G} the exact
determination of the rate at a specific metallicity is difficult. If
we restrict ourselves to nearby supernovae where host galaxies have
metallicities in the range from Solar to maybe half Solar, we find a
relative rate ratio of type Ib/c to type II supernovae or
0.29$\pm$0.17 \citep{2013MNRAS.436..774E,2015MNRAS.452.2597X}, where
again we are limited be large uncertainties. Despite these significant
uncertainties in any comparison single star models predict ratio below
this value while binary models predict a ratio close to or above this
value. An exact match is difficult because star-formation history and
metallicities distributions must be modelled of the stellar
population. But a stellar population with a high binary fraction
generally provides a better match to the observed ratio. This is
consistent with multiple equivalent studies over a significant period
of time
\citep[e.g.][]{1992ApJ...391..246P,1998A&A...333..557D}. Finally
recent work by \citet{2017A&A...601A..29Z} shows that progenitors
dominated by a population of interacting binary stars also reproduced
the observed core-collapse supernova delay time distribution better
than a single star only population.

\subsection{GW mergers}

In a massive binary system, if it can survive two supernovae and
remain bound then then two remnants, either neutron stars or black
holes, will slowly inspiral due to emission of gravitational radiation
\citep[e.g.][]{2016PhRvL.116x1102A,2017ApJ...848L..12A}. Since the
detection of the first merging black hole binary, GW150917, there have
been to date 5 pairs of merging black holes and one double neutron
star merger detected. To round off our holistic study of population
synthesis of massive stars, the same model that explains the observed
post-main sequence star ratios and supernova relative rates of
different types also needs to predict the mass range and rate of these
binaries. This is still work in progress and there are many binary
evolution codes now attempting to predict the rate of these mergers as
well as the masses of the observed merging objects. While this is
important, the same codes also need to make sure they correctly
predict the other, more visible aspects of stellar evolution that lead
up to formation of the remnants, rather than just the final ripples in
space-time from the merger of those remnants.

\section{Galaxies near and far}

The populations and distribution of stars within the Milky Way and
other galaxies in the local Universe reflects thirteen billion years
of evolution and complexity. While some of the stars, particularly the
massive stars, observed nearby formed in the locations where they
observed, others may have formed elsewhere within a gravitationally
bound system, or even in satellite stellar populations or progenitor
galaxies which merged over time into the current host. Most, if not
all, incorporate metals and other material which have been processed
by one or more earlier stellar populations and their resultant
supernovae.

A full understanding of the stellar populations nearby thus requires an
overview of processes which extend to cosmological scales in both time
and distance, including (but not limited to) the thermal history of
the Universe and variations in its chemical enrichment over cosmic
time.

Our primary information on these processes is derived from starlight,
sometimes reprocessed through dust or gas along the line of sight.
However for all but the most nearby galaxies, direct imaging and
resolution of individual stars is technically impossible
\citep[although note the existence of a claimed single-star,
  caustic-crossing lensing event in a distant $z=1.5$
  galaxy,][]{2016ATel.9097....1K}. The angular scales and sensitivity
required exceed the limits of even an ELT-class telescope. In a best
case scenario, it may be possible to measure the integrated light from
a single star forming region (down to $\sim$30\,pc in resolution)
making use of the spatial magnification associated with gravitational
lensing \citep[e.g.][]{2017ApJ...843L..21J}.  However in a more normal
case, resolving scales below 1\,kpc in galaxies is difficult from
space, and requires adaptive optics and/or exceptional conditions from
the ground. As a result, the emission detected represents the
integrated light of one or more stellar populations.  In many cases,
ultraviolet and optical light will be dominated by some combination of
the youngest and most massive stars, simply because these outshine
their more numerous but much fainter fellows.  These can be fit with a
simple stellar population, or with a straight-forward model such as
one continuously forming stars at a constant rate. At long
wavelengths, moving towards the infrared, or for galaxies dominated by
a massive underlying old stellar population, the contributions of
different elements of the stellar population, or from different stages
of the star formation history, can be more equal
\citep[e.g.][]{2003MNRAS.341...33K}.  In these cases, knowledge of
both the stellar mass function and the star formation history is
required, or needs to be derived from the available data.  In such
instances, the outputs of a stellar population and spectral synthesis
code are an essential tool for interpreting galaxy photometry and
spectroscopy.

This has been recognised for many years, and a number of spectral
synthesis models have been developed with the specific aim of
modelling the complex stellar populations seen in galaxies. These
include the very successful GALAXEV model set \citep[][and later
  references]{2003MNRAS.344.1000B} which combines theoretical stellar
evolution tracks with simple star formation histories and applies
empirical models for nebula emission at young stellar ages, and the
MAGPHYS models \citep{2008MNRAS.388.1595D} which aim primarily to
explore the non-stellar components of galaxies, extending into the
infrared and submillimetre, coupling these to the parent stellar
population through an energy balance between dust absorption and
re-emission. These and similar models have been used with great
success to explore the properties of galaxies in the nearby Universe,
where empirical calibration and multiple analysis methods can be used
to check their application and precision
\citep[e.g.][]{2004ApJ...613..898T,2004MNRAS.351.1151B,2011MNRAS.418.1587T}. However,
as observational samples of galaxies push to earlier times, when the
Universe was less metal rich, star formation occurred with a higher
specific rate density, and the stellar populations are significantly
younger \citep[see e.g.][for a recent review]{2014ARA&A..52..415M}, we
are moving into new regimes, where the efficacy of such models has not
been tested. What is more, the identification of extreme classes of
local galaxies which share similar physical conditions, and are thus
classed as analogues to galaxies in the distant Universe
\citep[e.g.][]{2005ApJ...619L..35H,2009MNRAS.399.1191C,2014A&A...568L...8A,2014MNRAS.439.2474S},
has demonstrated that even nearby galaxy-scale stellar systems can lie
well outside the normal range of properties predicted by the older
generation of population synthesis models.

At the same time, the same distant or extreme galaxy spectra present
an interesting opportunity to test and constrain stellar
modelling. The conditions which prevail in different regions of the
Universe, and at different epochs of cosmic history, are often rare in
the local Universe.  In particular the distant Universe (where the
galaxies we observe emitted their light within a few billion years
after the Big Bang) has attracted interest in this respect.  As
mentioned above, the stellar populations at those early times are
typically young ($<$1\,Gyr in age), and form from gas clouds with both a
lower overall metal enrichment (well below half-Solar), and
potentially very different abundance ratios to those common in the
local Universe (due to different enrichment processes).  While old
stellar populations such as Globular Clusters nearby tend to be
enhanced in alpha-process elements, as the result of a higher ratio of
core-collapse supernovae to other nucleosynthesis channels, at high
redshifts we see the same alpha-enhancement in much younger stellar
populations, dominated by more massive stars. At the same time
observations of galaxies in the distant Universe are weighted towards
the rest-frame ultraviolet spectral region, redshifted longwards of
the atmospheric cut-off and into the observed frame optical or
near-infrared.

Each of these conditions suggest that aspects of massive star
evolution, including their spectral evolution, lifetimes and metal
yields, are likely to be far more clearly manifest in distant galaxies
(and their local analogues) than in typical galaxies in the local
Universe. In this regime, binary interactions are expected to be important.
Thus while binary stellar population models are required to
interpret the light of these integrated stellar populations, the same
observations place a constraint on the models: any that fails to
reproduce the observed properties of star-forming galaxies given
plausible assumptions must be considered as suspect.

\section{Ionizing radiation fields and H\,II regions}

Early indications of the unusual stellar populations dominant in the
distant Universe came in the detection of nebular line emission,
particularly from lines requiring a hard ionising radiation spectrum
such as the rest-frame ultraviolet He\,II 1640 Angstrom emission line
from galaxies at $z>2$. These appeared to include both broad \citep[e.g.][]{2003ApJ...588...65S}
  and narrow \citep[e.g.][]{2010ApJ...719.1168E} components,
suggestive of strong stellar winds and intensely ionised nebular
regions respectively, while exhibiting no evidence for an AGN-powered
emission component. Studies of the He\,II feature have recently been
complemented by observations of sources with C\,III] and C\,IV in
  emission – again, with strong constraints indicating a likely
  stellar source for their photoionization \citep[e.g.][]{2015MNRAS.450.1846S,2015MNRAS.454.1393S,2017A&A...608A...4M}. 
  At the same time, galaxies at still earlier
  times ($z>5$) were exhibiting very blue ultraviolet colours
  \citep[e.g.][]{2005MNRAS.359.1184S,2010ApJ...708L..69B}. While
  there is certainly evolution in the dust extinction in the galaxy
  population, the slopes observed were suggestive of a much hotter
  stellar spectrum than typical at lower redshifts, and a resultant
  excess of hard ionising photons.

Early theoretical work suggesting that Population III
(i.e. essentially metal-free, primordial stars) were required to
explain these observations \citep[e.g.][]{2003A&A...397..527S,2006Natur.440..501J}
overlooked alternative sources of hard ionising photons, in particular
the strongly metal-dependant contribution of binary evolution
pathways, and the hot stars they produce. Initial work on the spectral
synthesis application of the BPASS binary stellar evolution models was
inspired, at least in part, by the desire to confront these
observations \citep[see][]{2009MNRAS.400.1019E,2012MNRAS.419..479E}. 
It demonstrated that
inclusion of binary evolution pathways was capable of generating
stronger ionising photon radiation fields, and provided a good fit to
a range of observations in both extreme local sources and to galaxies
in the distant Universe.

More recently, both the weight of evidence for physical conditions in
these systems and the models required to interpret them have developed
significantly. The advent of multi-object near-infrared spectrographs
on large telescopes (notably MOSFIRE on Keck), have rendered the
rest-frame optical accessible at $z>2$, and allowed for direct
comparison between the strong recombination line spectral diagnostics
seen in large local surveys such as the SDSS and those observed in
distant galaxies. One of the most striking results has been the
discovery of an offset in the Baldwin, Phillips and Terlevich [BPT, \citeyear{1981PASP...93....5B}] photoionization-sensitive diagnostic diagrams, cementing the
requirement for a hard ionizing radiation spectrum \citep[e.g.][]{2014ApJ...785..153M,2016ApJ...820...73H}. The same offset is seen in selected local
analogue galaxies, which have similarly high star formation densities \citep{2014MNRAS.444.3466S}.

Crucially, a number of these galaxies, while clearly consistent with a
star forming galaxy locus, lie above the `Maximal’ starburst line
defined by \cite{2001ApJ...556..121K}. This requires that their stellar
radiation field is harder than that derived from a combination of
single star spectral synthesis and nebular emission models. As a
number of studies have now shown \citep[e.g.][]{2014MNRAS.444.3466S,2016MNRAS.456..485S,2014ApJ...795..165S,2016ApJ...826..159S,2017ApJ...836..164S,2017A&A...608A..11G,xiao}
have now shown, binary spectral synthesis has two
important effects which tend to produce harder galaxy spectral
models. Firstly, binary interactions lead to the inclusion in
synthetic populations of stripped helium stars, and also stars which
undergo significant rotational mixing. The resultant higher surface
temperatures result in atmosphere models which emit a significant
fraction of their light shortwards of the Lyman limit, creating large
and hot photoionization regions.  Secondly, the range of timescales
for evolution of such stars extends the ionizing lifetime of a stellar
population beyond the few Myr required for the most massive stars to
live and die in a single star model. This extended lifetime increases
the fraction of galaxies and range of conditions in which large
photoionised regions are expected, and also boosts the contribution of
these to galaxies with continuous or ongoing starbursts.

Given the important role of metal opacities in driving stellar winds,
and the effect these have on stellar evolution, such binary effects
are strongly metallicity dependent. Consequently, the effects of
binary evolution are relatively slight in the bulk of highly
metal-enriched, slowly star-forming local galaxies, explaining the
success of single star models in fitting these. Nonetheless, an
analysis of star forming regions in local dwarf and spiral galaxies
suggests that BPASS binary models perform at least as well as single
star models in fitting their recombination line ratios, while yielding
slightly higher typical ages [Xiao et al, 2018, submitted].

The increasingly clear importance of binary population synthesis in
interpreting distant galaxies can largely be attributed to the effects
of cosmic metallicity evolution, together with a shift towards younger
stellar populations.

\section{Photon production, photon escape and reionization}

The hardness of the stellar radiation field is a key ingredient in
understanding the role of galaxies in a key phase change in the
thermal history of the Universe: the epoch of reionization. During
this period, the first luminous sources – most likely star-forming
galaxies – gradually ionised their immediate surroundings for the
first time since hydrogen atoms formed at $z\sim1100$. Individual
galaxy-scale H\,II regions expanded and overlapped over an extended
period, until the Universe was highly ionised. The topography of
ionised gas in this epoch should thus reflect both the distribution of
star forming galaxies, and the ionizing photon output of their stellar
populations. The Square Kilometer Array (SKA) will explore the power
spectrum of the epoch of reionization. Detailed mapping of the
ionisation balance intergalactic medium at this time is still well
beyond our capacity, and will likely remain so until the next
generation of H I mapping telescopes is designed and constructed.

Nonetheless, strong observational constraints on the reionization
transition do already exist. Signals imprinted on the cosmic microwave
background suggest that the Universe was 50\% ionised somewhere around
$z\sim9$. Evolution in the characteristics of Lyman-alpha line emission,
and the detection of ionised troughs in the absorption spectra of
distant sources (sometimes known as ‘cosmic lighthouses’), suggest
that the ionised fraction was rapidly rising between $z\sim6$ and $z\sim10$ \citep[see][and references therein]{2016A&A...594A..13P}, although line of
sight variations make a definitive timeline challenging to establish.

However reconciling this thermal history, observed in neutral
hydrogen, with the star formation history, observed in rest-frame
ultraviolet stellar emission, has proved challenging. Considerable
flexibility exists in estimates of parameters such as clumpiness
(and hence self-shielding) in the intergalactic medium, its density
and precise temperature. Nonetheless most estimates have suggested
that the observed galaxy population may struggle to produce sufficient
photons to reionize the Universe \citep[see e.g.][]{2015ApJ...802L..19R}.

\begin{figure}
\begin{center}
\includegraphics[width=0.8\textwidth]{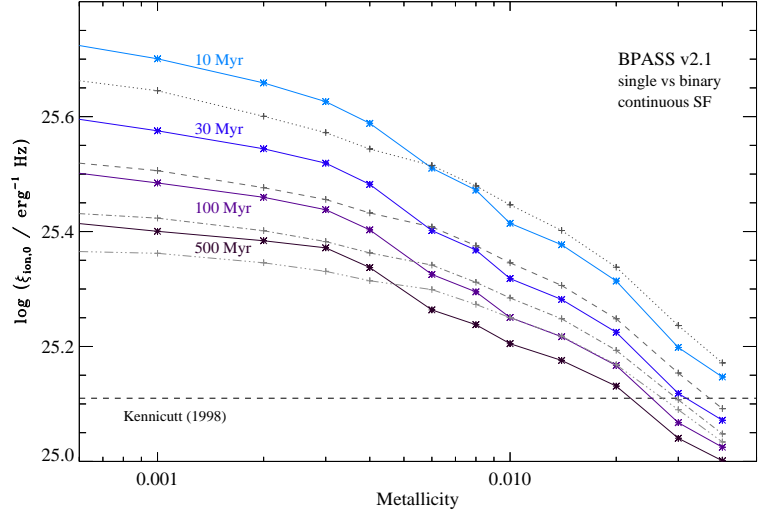}
\end{center}
\caption{The metallicity and stellar population age dependence of the
  ionizing photon production efficiency, $\xi_\mathrm{ion}$. This
  model-dependent parameter is crucial for placing observational
  constraints on the ionizing photon production during the
  reionization epoch. We show value for binary populations (solid,
  coloured lines) and single star populations (broken lines), at four
  different times after the onset of a constant, ongoing starburst
  event. Here we use the BPASS v2.1 models of
  \citet{2017PASA...34...58E}.\label{fig:xi_ion}}
\end{figure}

A key theoretical input into such estimates is the ionizing photon
production efficiency, $\xi_\mathrm{ion}$. This gives the rate of
ionizing photons (emitted shortwards of the Lyman limit) that will
arise from a population with a given rest-1500 Angstrom continuum
luminosity density. Since the former cannot be directly observed even
in the local Universe, the latter is used as an observable proxy for
ionisation, and $\xi_\mathrm{ion}$ is determined from theoretical
arguments, indirect measurements or appropriate population synthesis
models (or, more usually, a combination of all three). As figure
\ref{fig:xi_ion} demonstrates, the value of this parameter is
sensitive to both the stellar metallicity and age of an ongoing
continuous starburst. It is also sensitive to more complex star
formation history, and, needless to say, to the stellar population
synthesis model being used. In particular, the presence of binary
stars in a modelled population has a strong metallicity-dependent
effect, suppressing $\xi_\mathrm{ion}$ at metallicities close to Solar
and strongly boosting it at low metallicities.

Both the requirements of reconciling reionization timescales with
galaxy observations
\citep[e.g.][]{2016MNRAS.458L...6W,2016MNRAS.459.3614M,2016MNRAS.456..485S},
and the indirect measurements that can be obtained by considering the
reprocessed nebular emission spectrum in individual galaxies
\citep[e.g.][]{2015MNRAS.454.1393S,2017arXiv171100013S}, push the
required values of $\xi_\mathrm{ion}$ well above those suggested by
spectral synthesis models that neglect binary evolution effects, while
favouring the BPASS models, with their harder ionizing spectrum, and
resultant higher ionizing photon production efficiency.

\section{Looking forward} 

While significant progress has been made on studying the effects of
interacting binaries on the population and spectral synthesis there is
still much to do. Much of this resolves around extending the spectral
synthesis to the supernovae created by a binary population.

As discussed above the relative rate of different SN types can only be
reproduced by a stellar population including binaries, which increases the
number of stars that lose their hydrogen envelope. However an
important question is, what will those supernova actually look like?
Do they match the observed lightcurves and spectra of the SNe? To date
SN lightcurve models consider a few different models and only recently
have large numbers of models been computed.

What is required is a supernova population and spectral synthesis,
analogous to the stellar population and spectral synthesis. Groups are
now taking this models and using various open source supernova
simulation codes to predict the observed lightcurves of SNe. The
general result is that single stars the light curves differ only
slightly due to the different internal structure of the stellar
models, as they are broadly the same. However due to the variety of
progenitor structures that arise from binary interactions a much
greater range of possible lightcurves are possible.

Future work and synthesis like this are key, especially studies and models that
explore the full possible variety of binary evolution pathways. For
the first time we will have a prediction as to what the full variety
of possible stellar explosions may be, not only so we can match models
to observed explosions but also begin to understand what types of stellar
deaths we may be missing due to their faintness of rapid evolution.

Further work is also required to improve the connection between
modelling of simple stellar populations and that of entire
galaxies. In the latter, the effects of dust, diffuse interstellar
emission and complex star formation histories are often important and
difficult to disentangle or quantify. These can manifest in varied
ways, affecting not only galaxy photometry and line emission but also
less obvious properties, such as the chemical composition and
abundance patterns resulting from previous generations of supernovae
(commonly believed to lead to $\alpha$-element enhancement in old
stellar populations locally).  Nonetheless, distant galaxies provide
laboratories probing conditions atypical of the local Universe. There
is increasing evidence from observations of such sources for specific
photoionization phenomena that need to be explained. These appear
common in the unusual, metal poor and star-formation dense,
environments probed by young, distant galaxies, while remaining rare
in the resolved stellar populations studied locally.

Recent observations of distant galaxies suggest that the photospheric iron
abundance is significantly lower, relative to nebular oxygen
abundance, than assumed in common binary (and most other) models
\citep{2016ApJ...826..159S}. With the imminent launch of the {\em
  James Webb Space Telescope}, observations of photospheric line
blanketing in (rest-frame) far-ultraviolet spectra of distant galaxies
will become more straightforward and may be complemented by improved
measurements of both stellar and nebular abundance in the rest-frame
optical. Improved modelling, with a better understanding of the
abundance patterns and their effects on stellar evolution, will be
required to fully interpret such data.

Another example of an ongoing challenge is the strength of the He\,II
emission line observed in strongly star-forming galaxies. Observations
suggest that this is underestimated in models, which implies that
these underestimate the far-UV hard ionizing radiation field. Either a
change in the stellar population, for example a steeper IMF, or an
improved treatment of components such as accreting compact binaries
may be required to resolve this challenge, and must be accompanied by
an understanding of why such changes are required in the conditions
specific to the sources under observation.

\section{Summary}

In this chapter we have discussed some of the current findings in
population and spectral synthesis that show why including interacting
binaries is important. Also we have shown, by discussing a broad range
of observables, how by using a holistic view and considering
varied and different observations it is becoming possible to
firmly constrain the uncertainties of binary population synthesis and
increase the predictive power and usefulness of the many binary pop synth codes.

\bibliographystyle{plainnat}
\bibliography{ref}{}

 \copyrightline{} 
 \printindex
    
\end{document}